\documentclass[prc,superscriptaddress,showpacs,floatfix]{revtex4}
\usepackage[dvips]{graphicx}
\usepackage{epsfig}
\usepackage{pst-plot}
\usepackage{bm}
\usepackage{bbm}
\usepackage{mathrsfs}
\usepackage{amsfonts}
\usepackage{amssymb}
\usepackage{lscape}

\begin{document}

\title{Gamow shell model and  realistic nucleon-nucleon interactions}
\author{G.~Hagen}
\affiliation{Department of Physics and Astronomy, University of Tennessee, 
  Knoxville, Tennessee 37996, U.S.A.}
\affiliation{Physics Division, Oak Ridge National Laboratory, P.O. Box 2008, Oak Ridge, TN 37831, U.S.A.}  
\author{M.~Hjorth-Jensen}
\affiliation{Department of Physics and Center of Mathematics for Applications, 
University of Oslo, N-0316 Oslo, Norway}
\author{N.~Michel}
\affiliation{Department of Physics and Astronomy, University of Tennessee, 
  Knoxville, Tennessee 37996, U.S.A.}
\affiliation{Physics Division, Oak Ridge National Laboratory, P.O. Box 2008, Oak Ridge, TN 37831, U.S.A.}  

\date{\today}
\begin{abstract}
We present a  new and efficient method to obtain a Gamow shell-model
basis and matrix elements generated by realistic nucleon-nucleon
interactions. We derive a self-consistent Hartree-Fock potential
from the renormalized N$^3$LO interaction model. The corresponding  Gamow
one-body eigenstates
are generated in a plane wave basis in order to build a Gamow shell-model set of
basis states for the closed shell nuclei $^4$He and $^{16}$O.
We address also the  problem of representing a realistic nucleon-nucleon
interaction in a
two-particle Berggren basis in the laboratory frame.
To achieve this, an expansion of matrix elements of
the residual nucleon-nucleon interaction in
a finite set of harmonic oscillator wave functions is used.
We show that all loosely bound and
narrow resonant states converge surprisingly fast.
Even broad resonances in these
two-particle valence systems converge within a reasonable number of
harmonic oscillator functions.
Examples of $^6$He and $^{18}$O Gamow shell-model calculations using
$^4$He and $^{16}$O as closed shell cores are presented.
This procedure allows Gamow shell-model calculations to be performed with all
realistic nucleon-nucleon interactions and with either momentum or
position space representations
for the Gamow basis. 
Perspectives for nuclear structure
calculations of dripline nuclei are outlined.
\end{abstract}
\maketitle

\section{Introduction}
\label{sec:introduction}
A challenge in modern nuclear physics is the description of nuclei far from the valley of stability.
These nuclei exhibit unusual features such as very low particle-emission 
thresholds, halo densities and unbound ground states.
A proper understanding of the mechanisms underlying the formation of such nuclei 
is presently  a great challenge to nuclear theory, especially the case of two-neutron 
Borromean halos such as $^{6}$He and $^{11}$Li.
The theoretical description of such exotic nuclei cannot be worked out within standard models
because of the appearance of strong couplings to the continuum. 

The extreme clusterization of Borromean nuclei into an
ordinary core nucleus and a veil of halo nucleons  
has motivated few-body approaches such as the hyperspherical harmonic method 
and momentum space Faddeev equations to these nuclei \cite{zhuk}. 
However, the few-body modeling of Borromean and halo nuclei
is not completely satisfying as the treatment of core excitations and 
the anti-symmetrization between core and valence nucleons is therein approximate.

An \emph{ab initio} description
of these nuclei, taking into account all relevant degrees of freedom, would alleviate
the defects of such cluster approaches.
To achieve this, a reformulation 
of the shell model using a single-particle basis of bound, resonant and scattering states 
appears to be the most straightforward method. 
The Continuum shell model  \cite{vz2003,volya_PRL,volya2005,rotter1,rotter2} and the 
recently developed shell model embedded in the continuum (SMEC)
\cite{csm1,csm2,csm3,Jimmy_PRL} offer such a possibility.
In SMEC, two subspaces of bound/quasi-bound states and scattering states are introduced 
and their coupling taken into account following the techniques discussed in for example  Ref.~\cite{rotter1,rotter2}.
However, most calculations have been performed with only one-particle decay channels.
While the theoretical formulation of SMEC with two-particle decay channels has been formulated 
(see Ref.~\cite{Jimmy_PRL} with applications to two-proton radioactivity),
exact three-body asymptotics have never been applied numerically.
The very rapidly increasing complexity of SMEC with many-body decay channels is a hindrance
to the study of cluster-emitting systems, such as in particular Borromean nuclei.

The newly developed Gamow shell model
\cite{michel1, michel2, michel3, witek1, witek2, roberto, betan, betan2,hagen1, hagen2}
has proven to be a reliable tool in order to probe the structure of such nuclei.
This model unifies structure and reaction properties of nuclei, and most importantly allows for an exact treatment
of antisymmetry and has no limitation on the number of particles in the continuum.
It is then particularly well suited for the study of Borromean nuclei.
The starting point of the Gamow shell model is the Berggren completeness relation,  
where bound, resonant and scattering states are treated on an equal footing 
\cite{berggren,berggren1,berggren2,berggren3,lind}. 
The completeness relation is built upon bound, resonant states and 
an integral over a continuum of scattering states with complex energy.
This integral has to be discretized in order to be applied in numerical calculations.
A complete many-body Berggren basis is then constructed 
with Slater determinants integrating
bound, resonant and non-resonant discretized continuum orbitals.
The Gamow shell model can be seen as a direct generalization of the standard shell model, where
the standard harmonic oscillator  set of states is replaced by a Gamow basis.

An important question concerns the choice of the potential to generate the one-body Gamow basis states.
In Gamow shell-model calculations, the single-particle basis has normally been constructed from a Woods-Saxon or a 
Gaussian potential depicting  $^4$He or $^{16}$O cores, fitted to reproduce the single-particle states
of $^5$He and $^{17}$O, respectively \cite{michel1, hagen2}.
However, in a fully microscopic approach, the single-particle 
basis should be constructed from the free nucleon-nucleon interaction 
or more complicated three and/or many-body interactions. 
This can be done by summing various diagrams in many-body perturbation theory. At lowest order
this approach is given by the  Hartree-Fock approximation 
(see Ref.~\cite{michel1} where a Gamow-Hartree-Fock  basis was 
derived and applied to schematic interactions.)

In many-body perturbation theory, one cannot use the free nucleon-nucleon interaction,
since it yields strongly repulsive and/or diverging matrix elements at short
internucleonic distances. 
In order to remove these divergencies, renormalized nucleon-nucleon interactions have been constructed
from the Brueckner G-matrix approach \cite{bonatos,borromeo92,hko95}. 
The G-matrix is a soft interaction, which is obtained by resumming in-medium particle-particle
correlations. 

Recently, an alternative renormalization scheme 
which integrates out the high momentum components of the nucleon-nucleon
interactions has been proposed \cite{vlowk,suzuki4,suzuki5,bogner,nogga}. 
Using a similarity transformation of the two-nucleon Hamiltonian,
a Hermitian soft-core effective nucleon-nucleon interaction is obtained
in a model space defined by a cutoff $\Lambda$ in the relative momentum
between the nucleons. This renormalized interaction has become known as
a low-momentum nucleon-nucleon interaction, labeled $V_{\mathrm{low-k}}$. 
The interaction $V_{\mathrm{low-k}}$ is an energy and nucleus independent effective interaction
which reproduces nucleon-nucleon scattering data,
but displays a sizeable dependence on $\Lambda$.

In this work, the single-particle Gamow Hartree-Fock basis is constructed  using a renormalized 
interaction of the  $V_{\mathrm{low-k}}$ type, derived by similarity transformation 
techniques of the nucleon-nucleon interaction. 
Our renormalization scheme requires a plane wave basis formulation 
of the Schr\"odinger equation. Such a basis is a natural starting point since nucleon-nucleon interactions are 
usually derived explicitly in momentum space, as for example the N$^3$LO interaction \cite{n3lo1,n3lo2}. 
In order to perform Gamow Hartree-Fock calculations, the nucleon-nucleon 
interaction has to be defined by the coordinates of the  laboratory system. The transformation of the interaction 
from the relative and center of mass frame to the laboratory frame is performed with
the so-called vector brackets \cite{balian69,wc72,kkr79,bonatos}.
These are the less known momentum space analogs of the Moshinsky transformation coefficients 
of the harmonic oscillator representation,
generalizing the Talmi transformation to arbitrary bases.
In the presence of unbound states such as in a Gamow basis,
the single-particle potential has to be analytically continued in the complex 
$k$-plane. In Ref.~\cite{hagen1},
it was shown how a single-particle Berggren basis can be obtained by the 
contour deformation method in a basis of spherical Bessel functions.

For a microscopic approach to be fully consistent, 
the realistic nucleon-nucleon interaction should generate both a single-particle
basis through Hartree-Fock calculations and an effective nucleon-nucleon interaction to be diagonalized in the 
Gamow shell model. We can obtain this by letting
the renormalized nucleon-nucleon interaction to be expressed in a two-particle Berggren basis.
However, the difficulty in analytically continuing the vector transformation coefficients  to the complex $k$-space,
prevents such a derivation. In this work, an alternative approach to calculate
realistic interactions in Gamow bases is proposed.
The method is based on an expansion of the nucleon-nucleon interaction in a finite set of 
harmonic oscillator wave functions. Within this framework, the analytic continuation 
of the nuclear interaction is trivial, and matrix elements can therein be very efficiently calculated
through the use of the standard Talmi transformation. As will be shown, this method provides well converged energies 
and wave functions in the Gamow shell-model calculations with a small number of 
harmonic oscillator states.
In addition, as harmonic oscillator wave functions have a similar behavior in momentum and position space,
both momentum space and coordinate space representations can be used for the Gamow basis.
This method may also provide a solution to the problem of spurious center of mass motion in Gamow shell-model calculations.

The outline of the paper is a follows. 
In Sec.~\ref{sec:simtrans}, the derivation of a renormalized nucleon-nucleon 
interaction suitable for a perturbative many-body approach in the Gamow shell model is described.
In Sec.~\ref{sec:selfener}, the self-energy and Gamow Hartree-Fock single-particle basis
of the $V_{\mathrm{low-k}}$ interaction are constructed and applied to the $^4$He and $^{16}$O closed-shell nuclei.
Sec.~\ref{sec:vosc} outlines the harmonic oscillator expansion method for the nucleon-nucleon interaction,
and Sec.~\ref{sec:results} illustrates applications in Gamow shell-model calculations for two selected valence 
systems, $^{6}$He and $^{18}$O. There we discuss also 
the convergence of narrow and broad resonances as functions of the  
number of harmonic oscillator wave functions used in the expansion.
Sec.~\ref{sec:app2} points out the equivalence between the momentum and
the position space formulations of the Gamow shell model when the harmonic oscillator expansion method is used.
Finally, in Sec.~\ref{sec:conclusion} we outline our  conclusions and future 
perspectives.

\section{Renormalized nucleon-nucleon interaction}\label{sec:simtrans}

In order to build the Gamow Hartree-Fock potential in $k$-space 
and a Gamow shell-model Hamiltonian matrix, it is necessary to construct
the self-energy $\Sigma(k_al_aj_a,k_b)$ defined by the inclusion of various 
diagrams in many-body perturbation theory, discussed in Sec.~\ref{sec:selfener} .
Note here and in the following discussion the distinction between
$k_a,k_b$ and $k$ and $l_a$ and $l$. The notations $k_a$
or $l_a$ (latin letters) refer to the quantum numbers of a single-particle
state $a$, whereas $l$ or $k$ without subscripts (or with greek letters as subscripts) 
refer to the coordinates of the relative motion. 

To compute many-body perturbation diagrams, the nucleon-nucleon interaction
has to enter a perturbative treatment.
Hence, the free nucleon-nucleon interaction, giving rise to diverging matrix elements,
cannot be used directly and has to be renormalized.
Since parts of our formalism is based on computing the self-energy in a momentum basis,  it is
convenient here to use a renormalization scheme based on a cutoff in momentum
space as discussed by Bogner  {\em et al} \cite{vlowk} and 
Fujii {\em et al} \cite{suzuki4,suzuki5}.

This approach is based on two steps, a diagonalization in momentum space for relative momenta 
$k\in [0,\infty)$ 
of the two-body
Schr\"odinger equation and a similarity transformation \cite{suzuki4,suzuki5} to relative 
momenta $k\in [0,\Lambda]$, $\Lambda$ 
defining the relative momenta model space.
Typical values of $\Lambda$ are in the range of $\sim 2$ fm$^{-1}$.
The nucleon-nucleon interaction is diagonal in the center of mass motion. One can therefore 
easily map the full diagonalization problem onto a smaller space  via a similarity transformation
and obtain thereby an effective interaction for a model space defined for low momenta. This interaction
has been dubbed $V_{\mathrm{low-k}}$ in the literature, see for example Ref.~\cite{vlowk}.
The effective low-momentum interaction $V_{\mathrm{low-k}}$
is constructed in such a way that it reproduces 
exactly the main characteristics of the 
nucleon-nucleon wave function in the full space.

The interaction $V_{\mathrm{low-k}}$ looks attractive at first glance, 
but may generate undesirable features in the Gamow shell model.
Many-body calculations using a renormalized nucleon-nucleon interaction of the low-momentum 
type introduce a strong dependence on the cutoff $\Lambda $ in momentum space.
By integrating out high momentum modes of the nucleon-nucleon interaction,
one excludes certain intermediate excitations in the many-body problem. While the two-body 
problem is exact with $V_{\mathrm{low-k}}$, the three-body problem will not be.
In Ref.~\cite{nogga} $V_{\mathrm{low-k}}$  was accompanied with 
a $\Lambda$-dependent three-body force in order to reproduce the ground-state energies 
of $^4$He and $^3$H for each value of $\Lambda$. 
It is hoped that a three-body force is sufficient to eliminate the $\Lambda$ dependence for heavier nuclei.
However, if it turns out that one needs to go beyond three-body forces for nuclei with $A>4$, 
many-body calculations  starting with $V_{\mathrm{low-k}}$ are futile.
 

Alternatively, one could have 
defined a so-called $G$-matrix in momentum space as effective interaction \cite{hko95,bonatos,borromeo92,isaac98}. 
The latter introduces a dependence on the chosen starting energy 
and a reference to a given Fermi energy. This dependence can be eliminated by introducing
for example higher-order terms in many-body perturbation theory \cite{isaac98}.
We relegate such a discussion to future work. 
It must be stressed that the aim here is to demonstrate the feasibility
of obtaining a single-particle basis for Gamow shell-model calculations using a realistic interaction.
We adopt therefore a pragmatic approach and use $V_{\mathrm{low-k}}$ simply because it is easier to
implement numerically in order to renormalize the nucleon-nucleon interaction.

In the following  we outline the procedure to obtain a Hermitian 
interaction $V_{\mathrm{low-k}}$ based on the
similarity transformation discussed in Refs.~\cite{suzuki2,suzuki3,suzuki4,suzuki5}. 
A unitary transformation can be parametrized in terms of the model space $P$ and the excluded
space $Q$ via 
 \begin{equation}
 U = \left( \begin{array}{cc} P (1 + \omega ^\dagger \omega  )^{- 1/2}
 P & - P
 \omega ^\dagger ( 1 + \omega \omega ^\dagger )^{- 1/2}  Q \\
 Q \omega  ( 1 + \omega ^\dagger \omega  )^{- 1/2} P
 & Q (1 +  \omega \omega ^\dagger )^{- 1/2} Q
 \end{array} \right),
 \end{equation}
where the wave operator $\omega$ is defined to satisfy the condition
\begin{equation} 
\label{eq:decoup}
\omega  = Q \omega  P,
\end{equation}
 the so-called decoupling condition \cite{okubo}.
 Note that the unitary transformation 
 is by no means unique. In fact, one can construct infinitely many different 
 unitary transformations
 which decouple the $P$ and the $Q$ subspaces, as discussed by Kuo {\em et al} \cite{tom2004}.
 The above transformation  
 depends only on the operator $\omega$ which mixes the $P$ and $Q$
 subspaces and is in some sense ``the minimal possible'' 
 unitary transformation. 
 Following the method of Ref.~\cite{suzuki4}, one obtains
 \begin{equation}
 U=(1+\omega-\omega ^{\dagger})
 (1+\omega \omega ^{\dagger} +\omega ^{\dagger}\omega )^{-1/2}.
 \end{equation}
 The above operator $U$ leads to the effective interaction $\tilde{V}$ using the definition
 \begin{equation}
 \label{eq:V_eff}
 \tilde{V}=U^{-1}(T+V)U-T,
 \end{equation}
 where $T$ is the kinetic energy of the nucleons 
and $V$ is the free nucleon-nucleon interaction.

To express the renormalized interaction in momentum space, 
one starts with the Schr\"odinger equation for the relative momentum $k$, 
\begin{equation}
  \int dk'\:{k'}^2 \langle k\vert T + V \vert k'\rangle \langle {k'}\vert \psi_{\alpha}\rangle
  = E_{\alpha} \langle k\vert \psi_{\alpha}\rangle, 
\end{equation}
where the plane wave states are eigenfunctions of the kinetic energy 
operator $T$ in the relative system and form a complete set 
\begin{equation}
  \int dk \:k^2 \vert k\rangle \langle k\vert = \mathbbm{1}.
  \label{eq:bessel_complete}
\end{equation}
The momentum space Schr\"odinger equation is solved as a matrix equation
by discretizing the integration interval by some suitable rule, (here, the Gauss-Legendre quadrature is used). 
The discretized Schr\"odinger equation reads
\begin{equation}
  \sum_{\gamma} w_{\gamma} k_{\gamma}^2 \langle k_{\delta}\vert T+V
\vert k_{\gamma} \rangle \langle k_{\gamma}\vert\psi_{\alpha}\rangle 
  =  E_{\alpha} \langle k_{\delta}\vert \psi_{\alpha}\rangle,
  \label{eq:vlowk1}
\end{equation}
where $k_{\gamma}$ are the integration points and $w_{\gamma}$ the corresponding quadrature weights.
Introducing 
$\vert \bar{k_{\delta}} \rangle = k_{\delta}\sqrt{w_{\delta}} \vert k_{\delta} \rangle $, Eq.~(\ref{eq:vlowk1}) 
becomes
\begin{equation}
  \sum_{\gamma}  \langle \bar{k_{\delta}}\vert T+V \vert \bar{k_{\gamma}} \rangle 
  \langle \bar{k_{\gamma}}\vert\psi_{\alpha}\rangle 
  =  E_{\alpha} \langle \bar{k_{\delta}}\vert\psi_{\alpha}\rangle, 
  \label{eq:vlowk2}
\end{equation}
where 
\begin{equation}
  \sum_{\delta=1}^N \vert \bar{k_{\delta}} \rangle\langle \bar{k_{\delta}}\vert = \mathbbm{1}, \:\:
  \langle \bar{k_{\delta}}\vert \bar{k_{\gamma}} \rangle = \delta_{\delta,\gamma}.
\end{equation}
The matrix elements of the Hamiltonian are expressed in the plane wave basis :
\begin{equation}
  H_{\delta,\gamma} = \langle \bar{k_{\delta}}\vert T+V \vert \bar{k_{\gamma}} \rangle  = 
  {k_{\delta}^2\over m}\delta_{\delta,\gamma} + 
\sqrt{w_{\delta}w_{\gamma}}k_{\delta}k_{\gamma} V(k_{\delta},k_{\gamma}),
\end{equation} 
where $m$ is the average of the proton and the neutron masses.
The full space is now divided in a model space $P$ and an orthogonal complement 
space $Q$. The model space $P$ consists in the $N_P$  plane wave states
lying below the cutoff $\Lambda $, and the $Q$-space
consists of the remaining states, viz.
\begin{equation}
  P = \left\{ \vert \bar{k}\rangle, \:\: \vert k\vert  \leq \Lambda \right\}, \:\:
  Q = \left\{ \vert \bar{k}\rangle, \:\: \Lambda < \vert k\vert < \infty \right\}.
\end{equation}
The model space is thus defined for all momenta $k \in [0,\Lambda]$ fm$^{-1}$.
In order to obtain an effective interaction in the model space $P$,
the decoupling condition in Eq.~(\ref{eq:decoup}) 
has to be fulfilled.
Once the transformation matrix $\omega $ in the plane wave basis
$ \langle \bar{k}\vert Q\omega P\vert\bar{k}\rangle $ is obtained, the
low-momentum effective nucleon-nucleon interaction $V_{\mathrm{low-k}}$ reads
\begin{eqnarray}
  \nonumber \langle \bar{k}\vert P V_{\mathrm{low-k}} P\vert \bar{k}'\rangle & = & 
  \sum_{k''}\sum_{k'''}
  \langle \bar{k}\vert P(P+\omega^{\mathrm{T}}\omega )^{1/2} P\vert \bar{k}'' \rangle
  \langle \bar{k}''\vert P(T+V)P \vert \bar{k}'''\rangle 
  \langle \bar{k}'''\vert  P(P+\omega^{\mathrm{T}}\omega )^{-1/2}P \vert \bar{k}'\rangle \\
  & - & {k^2\over m} \delta_{kk'},
\end{eqnarray} 
see also Ref.~\cite{suzuki4} for further details. The effective model space interaction 
in the original plane wave basis $\vert k_{\delta} \rangle $ is then given by 
\begin{equation}
  \nonumber \langle {k_{\delta}}\vert V_{\mathrm{low-k}} \vert {k_{\gamma}}\rangle  =  
  { \nonumber \langle \bar{k_{\delta}}\vert V_{\mathrm{low-k}} \vert \bar{k_{\gamma}}\rangle 
    \over \sqrt{w_{\delta}w_{\gamma}} k_{\delta}k_{\gamma} },
\end{equation}
where $\left\{ \vert k_{\delta}\rangle, \:\vert k_{\gamma} \rangle\right\}  \in P$.

\section{Derivation of self-consistent Gamow Hartree-Fock basis}
\label{sec:selfener}

The renormalized nucleon-nucleon interaction $V_{\mathrm{low-k}}$ 
is defined in terms of various quantum numbers as follows
\begin{equation}
\left\langle klKL({\cal J})S T_z\right |
      V_{\mathrm{low-k}}\left | k'l'KL({\cal J})S T_z \right\rangle,
\end{equation}
where the variables $k$, $k'$ and $l$, $l'$
denote respectively relative and angular momenta,
while
$K$ and $L$ are the quantum numbers of the center of mass
motion. ${\cal J}$, $S$ and $T_z$ represent the total angular
momentum in the relative and center of mass system, spin and isospin projections,
respectively. 

The $A-$body Hamiltonian $H$ is defined as
\begin{equation}
  H = {1\over 2m} \sum_{i=1}^A {\bf k}_i^2 +  \sum_{i<j}^A V_{\mathrm{low-k}}(i,j).
\end{equation}
The spurious center of mass energy is removed by   writing the internal kinetic energy as
\begin{equation}
  T_{\mathrm{in}} = T - T_{\mathrm{c.m.}} = \left( 1 - {1\over A}\right) \sum_{i=1}^A 
  { {\bf k}_i^2\over 2m } - \sum_{i<j}^A { {\bf k}_i \cdot {\bf k}_j \over mA }. 
\end{equation}  
The introduction of an additional two-body term yields a modified two-body interaction
\begin{equation}
  H_{\mathrm{I}} = V_{\mathrm{low-k}} + V_{\mathrm{c.m.}}  = 
 \sum_{i<j}^A \left (V_{\mathrm{low-k}}(i,j)-\frac{{\bf k}_i \cdot {\bf k}_j}{mA}\right),
\label{eq:hmod} 
\end{equation}
resulting in a total Hamiltonian given by 
\begin{equation}
  H = \left( 1 - {1\over A}\right) \sum_{i=1}^A { {\bf k}_i^2\over 2m } + H_{\mathrm{I}}.
\end{equation}
In all calculations reported in this manuscript, the modified two-body Hamiltonian
defined in Eq.~(\ref{eq:hmod}) is employed. 

Starting from the renormalized momentum-space version of the 
nucleon-nucleon interaction $V_{\mathrm{low-k}}$, with matrix elements 
in the relative and center of mass system
system, one can obtain the corresponding matrix elements in the laboratory system
through appropriate transformation coefficients \cite{balian69,wc72,kkr79}. This transformation proceeds through the definition
of a two-particle state in the laboratory system.
With these coefficients,
the expression for a two-body wave function in momentum space
using the laboratory coordinates can be written as
\begin{widetext}
\begin{equation}
   \begin{array}{ll}
     &\\
     \left | (k_al_aj_at_{z_a})(k_bl_bj_bt_{z_b})JT_z\right \rangle =&
      {\displaystyle \sum_{lL\lambda S{\cal J}}}\int k^{2}dk\int K^{2}dK
      \left\{\begin{array}{ccc}
      l_a&l_b&\lambda\\\frac{1}{2}&\frac{1}{2}&S\\
      j_a&j_b&J\end{array}
      \right\}\\&\\
      &\times (-1)^{\lambda +{\cal J}-L-S}
      F\hat{{\cal J}}\hat{\lambda}^{2}
      \hat{j_{a}}\hat{j_{b}}\hat{S}
      \left\{\begin{array}{ccc}L&l&\lambda\\S&J&{\cal J}
      \end{array}\right\}\\&\\
      &\times \left\langle klKL| k_al_ak_bl_b\lambda\right\rangle
      \left | klKL({\cal J})SJT_z\right \rangle ,
   \end{array}
   \label{eq:relcm-lab}
\end{equation}
\end{widetext}
where $\left\langle klKL| k_al_ak_bl_b\lambda\right\rangle$
is the transformation coefficient (vector bracket) from the relative and center of mass system 
to the laboratory system  defined in Refs.~\cite{wc72,kkr79}.
The factor $F$ is defined as $F=(1-(-1)^{l+S+T_z})/\sqrt{2}$ if
we have identical particles only ($T_z=\pm 1$) and $F=1$ for 
different particles (protons and neutrons here, $T_z=0$).

The transformation coefficient $\left\langle klKL| k_al_ak_bl_b\right\rangle$ is 
given by \cite{balian69,wc72,kkr79}
\begin{equation}
\left\langle klKL| k_al_ak_bl_b\lambda\right\rangle=\frac{4\pi^2}{kKk_ak_b}\delta(w)\theta(1-x^2)A(x),
\label{eq:vectorbras}
\end{equation}
with
\begin{equation}
w= k^2+\frac{1}{4}K^2-\frac{1}{2}(k_a^2+k_b^2),
\end{equation}
\begin{equation}
x=(k_a^2-k^2-\frac{1}{4}K^2)/kK,
\end{equation}
and
\begin{equation}
A(x) = \frac{1}{2\lambda+1}\sum_{\mu}[Y^l(\hat{k})\times Y^L(\hat{K})]_{\mu}^{\lambda *}\times
[Y^{l_a}(\hat{k}_a)\times Y^{l_b}(\hat{k}_b)]_{\mu}^{\lambda}.
\end{equation}
The functions $Y^l$ are the standard spherical harmonics, $x$ is the cosine of the angle 
between $\vec{k}$ and $\vec{K}$ so that the step function takes input values from
0 to 1. In our codes, the coordinate system of Kuo {\em et al.}  \cite{kkr79} was chosen.
 
\begin{figure}
\resizebox{14cm}{8cm}{\epsfig{file=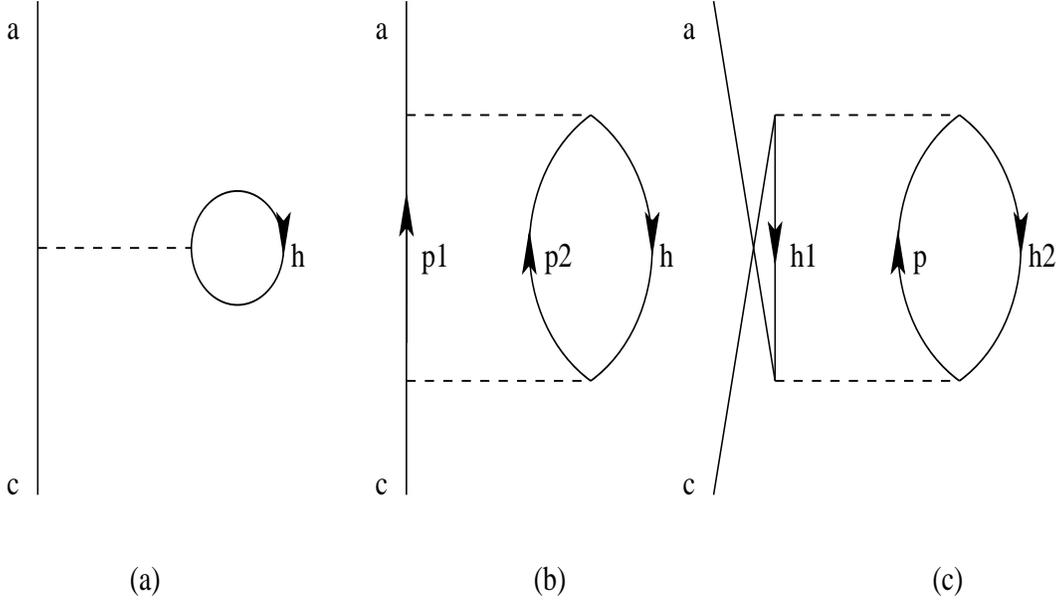}}
   \caption{Diagrams through second order of the interaction $H_{\mathrm{I}}$.
Diagram (a) is the first order Hartree-Fock term while  
diagrams (b) and (c) are the $2p-1h$ and  $2h-1p$ corrections, respectively.
The labels $a$ and $c$ represent the outgoing and incoming states, respectively.
The intermediate particle states are labeled as $p,p_1,p_2$ with upward arrows
and the intermediate hole states as $h,h_1,h_2$ with downward arrows.
The dotted lines represent the interaction $H_{\mathrm{I}}$.}
   \label{fig:figHF}
\end{figure}

To compute the Hartree-Fock (HF) diagram (see Fig.~\ref{fig:figHF} (a)) we need 
matrix elements in a mixed representation of bound and scattering states such as
\begin{equation}
   \left\langle (k_al_aj_at_{z_a})(n_bl_bj_bt_{z_b})JT_z\right |
    H_{\mathrm{I}}\left | (k_cl_cj_ct_{z_c})(n_dl_dj_dt_{z_d})JT_Z \right\rangle,
\end{equation}
where the labels $a$ and $c$ represent scattering states and $b$ and $d$ represent bound states.
All matrix elements discussed here are assumed to be 
antisymmetrized (AS).
Note that the two-body center of mass correction term $ V_{\mathrm{c.m.}} $ in Eq.~(\ref{eq:hmod})
is calculated directly 
in the laboratory coordinates using the Wigner-Eckart theorem, while the renormalized 
nucleon-nucleon interaction is given in laboratory coordinates using the transformation
given in Eq.~(\ref{eq:relcm-lab}).
The calculation of these matrix elements
requires the knowledge of two-body states in a mixed
representation with for example harmonic oscillator wave functions $R_{n_al_a}$
representing the bound states and plane waves for the resonant or continuum states
\begin{equation}
\hspace{-0.3cm}\left | (n_al_aj_at_{z_a})(k_bl_bj_bt_{z_b})JT_z \right \rangle=
    \int k_a^2dk_aR_{n_al_a}(k_a)
    \left | (k_al_aj_at_{z_a})(k_bl_bj_bt_{z_b})JT_z \right \rangle,
\end{equation}
where
$k_al_aj_a$ and $n_al_aj_a$ are respectively
plane wave and harmonic oscillator wave functions.
The two-body state is represented by the quantum numbers of the total angular momentum
$J$ and isospin projection $T_z$.

With these matrix elements, the expression for 
the HF diagram shown in Fig.~\ref{fig:figHF}(a) can be derived
\begin{widetext}
\begin{equation}
{\cal V}_{HF}(k_{a}k_{c}l_{a}j_{a}t_{z_{a}}) = \frac{1}{\hat{j_{a}}^2}
\sum_{J}\sum_{n_hl_hj_ht_{z_h}}\hat{J}^2
\left\langle
(k_{a}l_{a}j_{a}t_{z_a})
  (n_hl_hj_ht_{z_h})JT_z \right |
   H_{\mathrm{I}}\left | (k_{c}l_{a}j_{a}t_{z_{a}})
   (n_hl_hj_ht_{z_h})JT_z \right \rangle_{\mathrm{AS}},
\label{eq:hf}
\end{equation}
\end{widetext}
where $\hat{x}=\sqrt{2x+1}$ and $n_hl_hj_ht_{z_h}$ are the quantum numbers
of the nucleon hole states.
The variables
$l_{a}$, $j_{a}$, $t_{z_{a}}$ are the
orbital angular momentum, total angular momentum and isospin
projection ($t_{z_a}=\pm 1/2 $) of the incoming/outgoing
nucleon, and
$k_{a}$ ($k_{c}$) the outgoing (incoming)
particle momenta. 
The mixed matrix elements of the 
two-body center of mass correction term needed in the HF calculation reads
\begin{eqnarray}
  \nonumber
  \left\langle (k_al_aj_at_{z_a})(n_hl_hj_ht_{z_h})JT_z\right |
  -{ {\bf k}_1 \cdot {\bf k}_2 \over m A} \left | (k_cl_cj_ct_{z_c})(n_hl_hj_ht_{z_h})JT_z \right\rangle_{\mathrm{AS}} = \\
  \nonumber
      { 1 \over m A} \left( 2j_a+1 \right)\left( 2j_h+1\right) 
  \left(\begin{array}{ccc}j_a&1& j_h\\ 1/2&0& -1/2\end{array}\right)^2 
    \left\{\begin{array}{ccc}j_a&j_h& J\\ j_c&j_h& 1\end{array}\right\}\\
    \times\left\{ { 1+ (-1)^{l_a+l_h+1} }\over 2 \right\} k_ak_c R_{n_hl_h}(k_a) R_{n_hl_h}(k_c)
    \delta_{j_aj_c}\delta_{l_al_c}\delta_{t_{z_a}t_{z_h}} 
    \delta_{t_{z_c}t_{z_h}}. 
    \label{eq:pipj}
\end{eqnarray} 

In summary, if we limit ourselves to the computation of the HF contribution,
the expression for the self-energy reads
\begin{equation}
  \Sigma(j_{a}l_{a}k_{a}k_{c})={\cal V}_{HF}(j_{a}l_{a}k_{a}k_{c}).
\end{equation}
In this work, only the HF contribution is considered,
while in Ref.~\cite{borromeo92} 
the authors also studied contributions
from $2p-1h$ and $2h-1p$ intermediate states.
They yield an imaginary term which can be related to its real part via a 
dispersion relation.

To calculate the contributions from the $2p-1h$
diagrams like the example displayed in Fig.~\ref{fig:figHF} (b) (or 
similarly the $2h-1p$ diagram of Fig.~\ref{fig:figHF} (c))
we evaluate the imaginary part first. The real part is obtained
through the dispersion relations to be defined below.
The analytical expression for the imaginary contribution of the
$2p-1h$ diagram, which gives rise to an explicit energy dependence of
the self-energy, is
\begin{widetext}
\begin{eqnarray}
\lefteqn{{\cal W}_{2p-1h}(j_{a}l_{a}k_{a}
      k_{c}t_{z_{a}}\omega) \ = 
      {\displaystyle -\frac{1}
      {\hat{j_{a}}^2}\sum_{n_hl_hj_ht_{z_h}}
      \sum_{J}\sum_{lLS{\cal J}}\int k^{2}dk
      \int K^{2}dK\hat{J}\hat{T}}}\hspace{1cm}\nonumber \\
      &\times& \left\langle (k_{a}l_{a}j_{a}
      t_{z_{a}})(n_hl_hj_ht_{z_h})JT_z\right |
      H_{\mathrm{I}}\left | klKL({\cal J})SJT_z \right \rangle\nonumber \\
      &\times& \left\langle klKL({\cal J})SJT_z \right |H_{\mathrm{I}}
      \left | (k_{c}l_{a}j_{a}
      t_{z_{a}})(n_hl_hj_ht_{z_h})JT_z \right \rangle \nonumber
      \\ &\times& \pi\delta
      \left(\omega +\varepsilon_h -\frac{K^2}{4M_N} -
      \frac{k^2}{M_N} \right) \ ,
   \label{eq:2p-1h}
\end{eqnarray}
\end{widetext}
where $\omega$ is the energy of the incoming nucleon in a state $a$.
The quantities $klKL({\cal J})SJT_z$ are the
quantum numbers of the intermediate two-particle  state.
To compute the two-particle-one-hole diagram given by
the second-order diagram of Fig.~\ref{fig:figHF} (b), the following matrix elements 
are needed
\begin{equation}
   \left\langle (k_al_aj_at_{z_a})(n_bl_bj_bt_{z_b})JT_Z \right |
    H_{\mathrm{I}}\left | klKL({\cal J})ST_z \right\rangle .
\end{equation}
The  contributions to the real part of the self-energy from 
Eq.~(\ref{eq:2p-1h}) can be obtained through the following
dispersion relation
\begin{equation}
   {\cal V}_{2p-1h}(j_{a}l_{a}k_{a}k_{c}
   t_{z_{a}} \omega)=
   \frac{P}{\pi} \int_{-\infty}^{\infty}
   \frac{{\cal W}_{2p-1h}(j_{a}l_{a}k_{a}
    k_{c}
    t_{z_{a}} \omega')}{\omega'-\omega} d\omega',
    \label{eq:disprel}
\end{equation}
where $P$ takes the principal value of the integral. Since  ${\cal W}_{2p-1h}$
is different from zero only for positive values of
$\omega'$ and its diagonal matrix elements are negative,
this dispersion relation  implies that the diagonal elements of
${\cal V}_{2p-1h}$ will be attractive for
negative values of $\omega$. This attraction should increase for small
positive energies. It will eventually decrease and become repulsive
only
for large positive values of the energy of the interacting nucleon.
Similar expressions can also be derived for second-order diagrams with 
$2h-1p$ intermediate states \cite{borromeo92}.
Inclusion of these terms will be presented in a future work (see also the discussion in 
Sec.~\ref{sec:conclusion}).

The equations above for the nucleon self-energy are only valid along 
the real-energy axis. However, $ \Sigma(j_{a}l_{a}k_{a}k_{c}) $
has to be analytically continued from the real $k$-axis to
the complex $k$-plane in order to obtain a genuine Gamow shell-model single-particle basis.
The derivation of a Berggren basis in momentum space with the Contour Deformation Method was described in  Ref.~\cite{hagen1}. 
The analytical continuation of $ \Sigma(j_{a}l_{a}k_{a}k_{c})$ to the complex $k$-plane 
leads to theoretical and practical difficulties due to the appearance of Dirac and Heaviside distributions in 
vector brackets (see Eq.~(\ref{eq:vectorbras})). However, once the self-consistent self-energy
has been obtained along the real $k$-axis, the HF potential can be simply continued to the complex plane
via two sets of Fourier-Bessel transformations. 
To obtain a self-consistent HF potential in the complex $k$-plane, the following scheme is employed :
\begin{itemize}
\item The first step is to calculate ${\cal V}_{HF}$ self-consistently 
on a grid on the real momentum axis using the interaction  $H_{\mathrm{I}}$ of Eqs.~(\ref{eq:hmod}) 
and (\ref{eq:hf}).
\item When a self-consistent solution  has been obtained, ${\cal V}_{HF}$ is calculated in 
  position space via a double Fourier-Bessel transform.
  \begin{equation}
    {\cal V}_{HF}(jlrr')  =
    {2\over \pi } \int_{0}^{\infty}dk k^{2}\int_{0}^{\infty}dk {k}^{2} j_l(kr)j_l(k'r')
    {\cal V}_{HF}(jlkk').
    \label{eq:hf_coord}
  \end{equation}
\item Having obtained ${\cal V}_{HF}$ in the $r$-plane we may go back to the complex $k$-plane
  using one more Fourier-Bessel transformation.
  On an inversion symmetric contour $L^+$ in the complex $k$-plane, 
  the HF potential becomes
  \begin{equation}
    {\cal V}_{HF}(jlkk') =  
    \int_{0}^{\infty}dr r^{2}\int_{0}^{\infty}dr' {r'}^{2} j_{l}(kr)j_{l}(k'r')
  {\cal V}_{HF}(jlrr'),
  \end{equation}
\end{itemize}
where $k$ and $k'$ belong to the contour $L^+$ and are therefore complex.
The analytically continued single-particle Schr\"odinger equation on a general 
inversion symmetric contour then takes the form
\begin{equation}
\label{eq:neweq1}
{\hbar^{2}\over m_{\mathrm{eff}}}k^2\psi_{nlj}(k) + \int_{L^{+}} 
dk' {k'}^{2}{\cal V}_{HF}(j l kk')\psi_{nl}(k') = E_{nl}\psi_{nl}(k),
\end{equation}
with $m_{\mathrm{eff}} = 2m( 1 - 1/A)^{-1} $. 
Here both $k$ and $k'$ are defined on an inversion symmetric contour $L^+$ in the lower
half complex $k$-plane, resulting in  a closed integral equation. In order to solve this 
equation, the integral has to be discretized, and we finally end up with a complex symmetric 
matrix diagonalization problem, in analogy with Eq.~(\ref{eq:vlowk1}). 
This procedure results in a self-consistent Gamow Hartree-Fock basis, which is complete within the
discretization space, and includes bound, resonant and a finite set of 
non-resonant continuum states (see Ref.~\cite{hagen1} for details about one-body Berggren completeness relations in momentum space).

Based on this approach, the single-particle Gamow Hartree-Fock states of the closed-shell $^4$He and $^{16}$O are calculated
with a low-momentum nucleon-nucleon interaction constructed from the realistic 
N$^3$LO nucleon-nucleon interaction \cite{n3lo1,n3lo2}. 
For $^{16}$O we used the models spaces defined by 
$\Lambda =$  1.9, 2.0 and 2.1 $\mathrm{fm}^{-1}$, and for $^4$He the 
models spaces defined by  $\Lambda =$  1.8, 1.9 and 2.0 $\mathrm{fm}^{-1}$

\begin{figure}
\resizebox{8cm}{5cm}{\epsfig{file=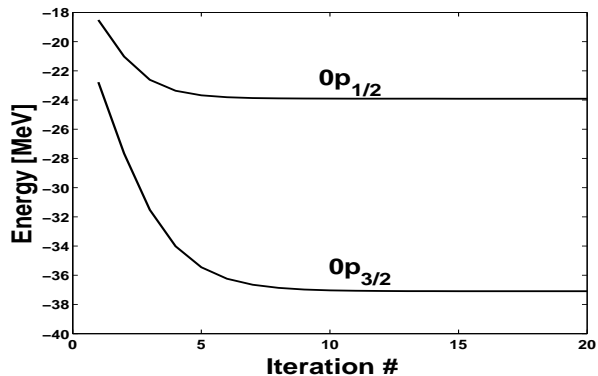} }
\caption{Convergence of the 
  $0p_{3/2}$ and the $0p_{1/2}$ energies in $^{16}$O with respect to iteration number 
  in the self-consistent HF calculation. Here we used a model space 
$\Lambda = 1.9 \mathrm{fm}^{-1}$ in the construction of $V_{\mathrm{low-k}}$.}
\label{fig:O16_sc} 
\end{figure}

Table \ref{tab:n3lo_sp_016} presents the neutron single-particle energies obtained
with a $^{16}$O core. We note that holes states are in general overbound for all chosen 
cutoffs $\Lambda$ 
and that the spin-orbit splitting 
between the $p_{3/2}$ and $p_{1/2}$ states is too large. The particle states are in better agreement with data
and for all model spaces the $d_{3/2}$ state is the only one which comes out as a resonance, in agreement with experiment. 
The spin-orbit spacing the $d_{5/2}$ and $d_{3/2}$ states is also fairly well reproduced. 
Fig.~\ref{fig:O16_sc} shows the convergence of the 
$0p_{3/2}$ and the $0p_{1/2}$ spin-orbit partners in $^{16}$O with respect to the number of iterations  
in the self-consistent HF calculation.
\begin{table}[htbp]
\caption{HF calculation of single-particle energies in $^{16}$O using 
  the low-momentum N$^3$LO nucleon-nucleon interaction for three
  different model spaces.
  The single-particle energies $E$ are given in MeV for both real and imaginary parts. 
  Experimental data are from Ref.~\cite{Fire}.}
  \begin{tabular}{rrrrrrrrr}
    \hline
    \multicolumn{1}{c}{} & \multicolumn{2}{c}{$\Lambda = 1.9$ $\mathrm{fm}^{-1}$}
    & \multicolumn{2}{c}{$\Lambda = 2$ $\mathrm{fm}^{-1}$}  
    & \multicolumn{2}{c}{$\Lambda = 2.1$ $\mathrm{fm}^{-1}$}  
    & \multicolumn{1}{c}{Expt.}\\
    \hline
    \multicolumn{1}{c}{$lj$}&\multicolumn{1}{c}{Re[E]}&\multicolumn{1}{c}{Im[E]} &
    \multicolumn{1}{c}{Re[E]}&\multicolumn{1}{c}{Im[E]}&
    \multicolumn{1}{c}{Re[E]}&\multicolumn{1}{c}{Im[E]} &
    \multicolumn{1}{c}{Re[E]}&\multicolumn{1}{c}{Im[E]} \\
    \hline
    $s_{1/2}$ &  -73.977 & 0.000 & -68.496 &  0.000 & -63.078 &  0.000 &  -44.000 & 0.000  \\
    $p_{3/2}$ &  -37.082 & 0.000 & -33.824 &  0.000 & -30.650 &  0.000 &  -21.840 & 0.000  \\
    $p_{1/2}$ &  -23.981 & 0.000 & -21.749 &  0.000 & -19.680 &  0.000 &  -15.664 & 0.000  \\
    $d_{5/2}$ &   -5.060 & 0.000 &  -3.810 &  0.000 &  -2.541 &  0.000 &  -4.143  & 0.000 \\
    $s_{1/2}$ &   -3.531 & 0.000 &  -2.556 &  0.000 &  -1.781 &  0.000 &  -3.273  & 0.000 \\
    $d_{3/2}$ &    5.189 &-1.669 &   5.353 & -1.928 &   5.419 & -2.155  &  0.937  & -0.048 \\ 
    \hline 
  \end{tabular}
  \label{tab:n3lo_sp_016}
\end{table}

Table \ref{tab:n3lo_sp_he5} presents the calculated self-consistent neutron 
single-particle energies with respect to a $^4$He closed shell core.
The calculated values for the $p_{3/2}$ and the $p_{1/2}$ energies are not so far from 
the experimental values. For a model space  $\Lambda  =1.8$ fm$^{-1}$,   
the obtained width of the $p_{3/2}$ resonance coincides 
with the experimental width $ 0.648 $ MeV, while
the calculated width of the $p_{1/2}$ resonance  ($\sim 7.4$ MeV) is 
larger than the experimental value $ 5.57$ MeV. The spin-orbit splitting
between  the $p_{3/2}$ and the $p_{1/2}$ levels is fairly well reproduced, the experimental value
is 1.27 MeV \cite{tilley} while our values vary from 1.36 to 1.68 MeV. 
Noting that our calculations are done at the Hartree-Fock level, there is clearly room
for improvements. However, although our results for hole states are overbound, 
we obtain a qualitatively correct spectrum.
\begin{table}[htbp]
\caption{Same as in Tab.~\ref{tab:n3lo_sp_016} for $^4$He. Experimental data from
Ref.~\cite{tilley}.}
\begin{tabular}{rrrrrrrrr}\hline
  \multicolumn{1}{c}{} &\multicolumn{2}{c}{$\Lambda = 1.8\mathrm{fm}^{-1}$}&
  \multicolumn{2}{c}{$\Lambda = 1.9\mathrm{fm}^{-1}$}&
  \multicolumn{2}{c}{$\Lambda = 2.0\mathrm{fm}^{-1}$}&
  \multicolumn{1}{c}{Expt.}  
  \\
  \hline
  \multicolumn{1}{c}{$lj$}&\multicolumn{1}{c}{Re[E]}&\multicolumn{1}{c}{Im[E]} &
  \multicolumn{1}{c}{Re[E]}&\multicolumn{1}{c}{Im[E]} &
  \multicolumn{1}{c}{Re[E]}&\multicolumn{1}{c}{Im[E]} &
  \multicolumn{1}{c}{Re[E]}&\multicolumn{1}{c}{Im[E]} \\
  \hline
  $ s_{1/2}$ &  -25.731  &   0.000  & -24.541 &  0.000 & -23.079 &  0.000 &  -20.578 & 0.000    \\
  $ p_{3/2}$ &    0.819  &  -0.325  &   1.041 & -0.479 &   1.287 & -0.667 &  0.890 &  -0.324 \\ 
  $ p_{1/2}$ &    2.497  &  -3.697  &   2.514 & -3.777 &   2.648 & -4.029 &  2.160 &  -2.785  \\
  \hline
\end{tabular}
\label{tab:n3lo_sp_he5}
\end{table}
Higher-order corrections should improve the agreement with experiment,
since $2p-1h$ and $2h-1p$ correlations provide additional
binding and improved spin-orbit splittings for particle states.

Many-body calculations using a renormalized nucleon-nucleon interaction of the low-momentum 
type will unfortunately introduce a model space dependence in momentum space (see 
Tables \ref{tab:n3lo_sp_016} and \ref{tab:n3lo_sp_he5}).
The model space dependence can only be eliminated by introducing the corresponding 
three- and many-body forces which the low-momentum two-body interaction induces. In Gamow shell-model
calculations, the inclusion of three-body forces is not feasible for the moment.
If one wants to minimize the effect from many-body forces, 
a better approach might be to use a $G$-matrix for nuclear matter as in Refs.~\cite{borromeo92,isaac98}.
However, higher-order correlations such as $2p-1h$
or $2h-1p$ contributions are necessary in order to minimize the dependence on the 
starting energy and the chosen Fermi energy \cite{isaac98}.
Whether the model space 
dependence of $V_{\mathrm{low-k}}$ can be softened by the inclusion of higher-order correlations such as $2p-1h$ and 
$2h-1p$ will be investigated in a future work.
Furthermore, although the $G$-matrix carries an explicit starting energy dependence, this dependence is needed
when one wants to compute for example spectral functions.

\section{Matrix elements of realistic nucleon-nucleon interaction with a  Gamow Hartree-Fock basis}
\label{sec:vosc}
As discussed in the previous section, 
starting from  matrix elements of the nucleon-nucleon interaction 
in the relative and center of mass system
system, one can obtain the corresponding matrix elements in the laboratory system
through appropriate transformation coefficients, see for example 
Refs.~\cite{balian69,wc72,kkr79} and the discussion in the previous section. 
This transformation proceeds through the definition
of a two-particle state in the laboratory system using the vector bracket transformation.
However, the latter is very complicated to 
handle in practical calculations beyond the Hartree-Fock level. One has also to face the problem 
of two-particle intermediate states not orthogonal to the 
incoming and outgoing states (see Eq.~(\ref{eq:2p-1h})).  
Another method is then needed in order 
to efficiently calculate effective nucleon-nucleon matrix elements in 
the complex $k$-plane and laboratory frame with a Gamow shell-model basis.

The renormalized nucleon-nucleon interaction in an arbitrary two-particle basis in the laboratory frame is given by
    \begin{equation}
     \langle ab \vert V_{\mathrm{low-k}} \vert cd \rangle = 
    \left\langle (n_al_aj_at_{z_a})(n_bl_bj_bt_{z_b})JT_z \right | V_{\mathrm{low-k}}
    \left | (n_cl_cj_ct_{z_c})(n_dl_dj_dt_{z_d})JT_z \right \rangle. 
    \label{eq:nn_lab_compressed}
  \end{equation}
The two-body state $\vert a b\rangle $ is implicitly coupled 
to good angular momentum $J$. The labels $n_{a...d}$ number all
bound, resonant and discretized scattering states with orbital and angular momenta ($l_{a...d},j_{a...d}$).

In order to efficiently calculate the matrix elements of Eq.~(\ref{eq:nn_lab_compressed}), we introduce a two-particle 
harmonic oscillator basis completeness relation
\begin{equation}
  \sum_{\alpha \leq \beta } \vert \alpha \beta\rangle \langle \alpha \beta \vert = \mathbbm{1}, 
  \label{eq:osc_completeness}
\end{equation}
where the sum is not restricted in the neutron-proton case. We introduce the greek single 
particle labels $\alpha , \beta $ for the single-particle harmonic oscillator  
states in order to distinguish them from the latin single-particle labels $a,b$ referring to Gamow states.
The interaction can then be expressed in the complete 
basis of Eq.~(\ref{eq:osc_completeness})
\begin{equation}
  V_{\mathrm{osc}} = \sum_{\alpha \leq \beta}\sum_{\gamma \leq \delta} 
  \vert \alpha \beta\rangle \langle \alpha\beta \vert  V_{\mathrm{low-k}} \vert \gamma \delta \rangle 
  \langle \gamma \delta \vert, 
  \label{eq:vosc}
\end{equation}
where the sums over two-particle harmonic oscillator states are infinite. The expansion coefficients
\begin{equation}
    \nonumber
    \langle \alpha\beta \vert V_{\mathrm{low-k}} \vert \gamma \delta \rangle  = 
    \left\langle (n_\alpha l_\alpha j_\alpha t_{z_\alpha})(n_\beta l_\beta j_\beta t_{z_\beta})JT_z 
    \right |V_{\mathrm{low-k}}
    \left | (n_\gamma l_\gamma j_\gamma t_{z_\gamma})(n_\delta l_\delta j_\delta t_{z_\delta})JT_z 
    \right \rangle,
    \label{eq:nn_lab_HO}
 \end{equation}
represent the nucleon-nucleon interaction in an antisymmetrized 
two-particle harmonic oscillator basis, and may easily be calculated using the 
well known Moshinsky transformation coefficients, see for example Ref.~\cite{lawson} for expressions. 

Matrix elements of Eq.~(\ref{eq:nn_lab_compressed}) are calculated numerically  
in an arbitrary two-particle Gamow basis by truncating the completeness 
expansion of Eq.~(\ref{eq:vosc}) up to $N$ harmonic oscillator two-body states
\begin{equation}
  \langle ab \vert V_{\mathrm{osc}} \vert cd \rangle \approx \sum_{\alpha \leq \beta}^N \sum_{\gamma \leq \delta}^N 
  \langle ab \vert \alpha \beta\rangle \langle \alpha\beta \vert  V_{\mathrm{low-k}} \vert \gamma \delta \rangle 
  \langle \gamma \delta \vert c d \rangle.
  \label{eq:nn_lab_approx}
\end{equation}
The two-particle overlap integrals $ \langle ab \vert \alpha \beta\rangle $ read
\begin{equation}
  \langle ab \vert \alpha \beta\rangle = 
	  { \langle a \vert \alpha \rangle\langle b\vert \beta \rangle
	    -(-1)^{J-j_\alpha -j_\beta} 
	  \langle a \vert \beta \rangle\langle b\vert \alpha \rangle 
	    \over \sqrt{ (1+\delta_{ab})(1+\delta_{\alpha\beta})} } 
\end{equation}
for identical particles (proton-proton or neutron-neutron states) and 
\begin{equation}
  \langle ab \vert \alpha \beta\rangle = \langle a \vert \alpha \rangle\langle b\vert \beta \rangle
\end{equation}
for the proton-neutron case.
The one-body overlaps $\langle a \vert \alpha \rangle $ are given by, 
\begin{equation}
  \langle a \vert \alpha \rangle = \int d \tau \: \tau^2 \varphi_a(\tau) R_\alpha(\tau)\:
  \delta_{l_al_\alpha} \delta_{j_aj_\alpha}\delta_{t_at_\alpha},
  \label{eq:sp_overlap}
\end{equation}
where $\varphi_a$ is a Gamow single-particle state and 
$R_\alpha$ is a harmonic oscillator wave function. The radial integral is either evaluated in 
momentum or position space, indicated by the integration variable $\tau$. 
The important point to notice is that the only numerical calculations
involving Gamow states are the overlaps of Eq.~(\ref{eq:sp_overlap}).
Hence, this expansion provides a simple analytical continuation of the nuclear interaction 
in the complex $k$-plane. More precisely, the expansion coefficients of Eq.~(\ref{eq:nn_lab_HO})
can always be calculated with real harmonic oscillator wave functions, and in the case of Gamow functions
spanning a complex contour $L^+$ (as in the momentum space representation), it is only the
overlap integrals of Eq.~(\ref{eq:sp_overlap}) which are analytically continued in the 
complex plane. These one-body overlap integrals converge in all regions of the complex 
plane which are of physical importance, due to the gaussian fall off of the harmonic oscillator wave functions.

The convergence with the number of harmonic oscillator states $N$ of the 
nuclear interaction expansion of Eq.~(\ref{eq:nn_lab_HO})
can however not be checked by considering the matrix elements of Eq.~(\ref{eq:nn_lab_approx})
when $N \rightarrow +\infty$. Indeed, they generally diverge when $N \rightarrow +\infty$,
due to the  long-range character  of the nuclear interaction in laboratory coordinates.
This reflects the fact that the convergence of the Gamow shell-model Hamiltonian with $N$ is \textit{weak},
because the representation of a Hamiltonian such as $V_{\mathrm{low-k}}$ 
in terms of a continuous Gamow basis is a distribution.
Actually, only eigenvalues and eigenfunctions of the Gamow shell-model Hamiltonian converge with $N$, 
which will be shown in particular cases in Sec.~\ref{sec:results}.

\section{Gamow shell-model  calculations of $^{6}$He and $^{18}$O.}
\label{sec:results}
\begin{figure}
  \resizebox{8cm}{5cm}{\epsfig{file=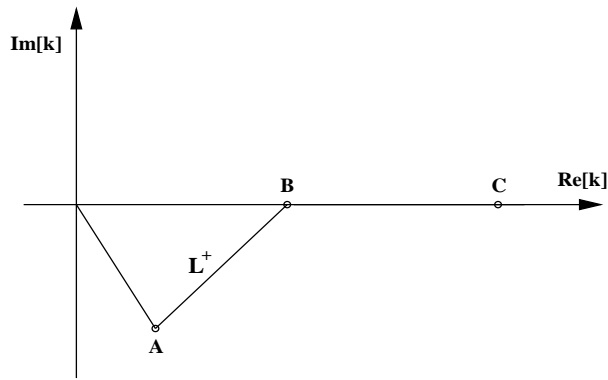}}
   \caption{Contour $L^+$ in the complex $k$-plane 
     used in construction of the single-particle Berggren basis. 
     The contour in specified by the points $A,B$ and $C$ discussed in the text.}
\label{fig:contour1} 
\end{figure}

In Sec.~\ref{sec:selfener} we constructed a self-consistent single-particle Gamow Hartree-Fock   
basis
starting from a realistic renormalized nucleon-nucleon interaction.
Ultimately this basis should be used to construct the 
effective interaction in a given model space and 
diagonalize the shell-model Hamiltonian.
Using the harmonic oscillator   expansion method of the nucleon-nucleon interaction,
we have a practical way of constructing the effective interaction to be incorporated in Gamow shell-model calculations.
It is our aim to investigate whether a finite 
truncation of the harmonic oscillator expansion given in Eq.~(\ref{eq:nn_lab_approx})
may yield converged energies and wave functions in Gamow shell-model 
calculations. 
As a first application we consider two nucleons moving outside a closed core.
The Gamow shell-model  Hamiltonian used reads 
\begin{equation}
H(1,2) = h_{\mathrm{HF}}(1) + h_{\mathrm{HF}}(2) + V_{\mathrm{eff}}(1,2),
\end{equation}
where $h_{\mathrm{HF}}$ is the self-consistently derived single-particle 
Hartree-Fock potential and $  V_{\mathrm{eff}}(1,2) $ is the effective 
interaction acting between valence particles.
In this work, all calculations are implemented up to first order in many-body perturbation theory,
where the effective interaction is defined as 
\begin{equation}
  V_{\mathrm{eff}}(1,2) = V_{\mathrm{low-k}}(1,2)-\frac{{\bf k}_1\cdot {\bf k}_2}{mA}.
\label{eq:hmod2} 
\end{equation}
It should be noted that we utilize the harmonic oscillator expansion method only for 
the nucleon-nucleon interaction part, i.e. $ V_{\mathrm{low-k}}(1,2) \approx V_{\mathrm{osc}}$,
while the center of mass correction term is treated exactly in the laboratory frame.  
The matrix elements of the center of mass term are given by
\begin{eqnarray}
  \nonumber
  \left\langle (n_al_aj_at_{z_a})(n_bl_bj_bt_{z_b})JT_z \right | {\bf k}_1\cdot {\bf k}_2 
  \left | (n_cl_cj_ct_{z_c})(n_dl_dj_dt_{z_d})JT_z \right \rangle = \\
  -\hbar^2 (-1)^{j_c+j_a+J}\left\{\begin{array}{ccc}j_a&j_b& J\\ j_d&j_c& 1\end{array}\right\}
  \langle n_al_aj_a \vert\vert \nabla_1 \vert\vert n_cl_cj_c\rangle
  \langle n_bl_bj_b \vert\vert \nabla_2 \vert\vert n_dl_dj_d\rangle, 
\end{eqnarray}
where the expression for the reduced matrix elements can be found in 
Ref.~\cite{lawson}. 
In our Gamow shell-model study of the two-particle valence systems with realistic interactions, $^6$He and $^{18}$O, 
a two-particle model space built  from the 
$s_{1/2}$, $p_{3/2}$, $p_{1/2}$, $d_{5/2}$ and $d_{3/2}$ single-particle states is used. Each combination
of the quantum numbers $lj$  
consists of 25 single-particle orbitals, totalling 115 orbitals. 
The same integration contour $L^+$  in the complex $k$-plane is used for all partial waves 
(see Fig.~\ref{fig:contour1}). In our $^6$He 
calculations
the contour is defined with $A = 0.28-0.12i$ $\mathrm{fm}^{-1}, B= 0.5$ $\mathrm{fm}^{-1} $ 
and $ C= 4$ $\mathrm{fm}^{-1} $,  
and in our $^{18}$O calculations
it is defined with $A = 0.52-0.12i$ $\mathrm{fm}^{-1}, B= 0.64$ $\mathrm{fm}^{-1} $ 
and $ C= 4$ $\mathrm{fm}^{-1} $.
Using Gauss-Legendre quadrature, the discretization of
$L^+$ has been carried out with $5$ points in the interval $(0,A)$, $7$ points in the
interval $(A,B)$ and 13 points in the interval $(B,C)$.
Convergence of the two-particle states with respect to the number 
of integration points has been checked. Our discretization of $L^+$ yields a
precision of the energy calculation better than $0.1$ keV for all states considered.
In the following discussion a model space defined by $\Lambda = 1.9$ $\mathrm{fm}^{-1}$ is employed.
The oscillator length is  fixed at $b=2$ fm. 

\subsection{ $^6$He results }
Table \ref{tab:he6_results} gives the convergence of the $0_1^+$ ground state and 
the $2_1^+$ excited state of $^6$He as functions of  increasing number of nodes in the harmonic oscillator 
expansion of the interaction.  A remarkable observation is that the 
$0_1^+$ ground and the $2_1^+$ excited states of $^6$He 
converge rather fast with respect to the number of harmonic oscillator
functions, since $n_{\mathrm{max}} = 10$ is sufficient to reach convergence. 
\begin{table}[htbp]
\caption{Convergence of the ${0_1}^+$ and the $ {2_1}^+ $ energies in ${}^6$He as functions of  
the number of harmonic oscillator nodes in the expansion of the realistic low-momentum 
N$^3$LO nucleon-nucleon interaction. The model space is 
given by $\Lambda = 1.9$ $\mathrm{fm}^{-1}$.
The harmonic oscillator length is chosen at $b=2$ fm. Energies are given in units of MeV.}
\begin{tabular}{rrrrr}\hline
  \multicolumn{1}{c}{} & 
  \multicolumn{2}{c}{$J^\pi = {0_1}^+$} & 
  \multicolumn{2}{c}{$J^\pi = {2_1}^+ $}\\
  \hline
  \multicolumn{1}{c}{$n_{\mathrm{max}}$} & 
  \multicolumn{1}{c}{Re[E]} & \multicolumn{1}{c}{Im[E] }& 
  \multicolumn{1}{c}{Re[E]} & \multicolumn{1}{c}{Im[E] }\\
  \hline
  4 &  -0.4760 &  0.0000 & 0.9504 & -0.0467  \\
  6 &  -0.4714 &  0.0000 & 0.9546 & -0.0461  \\
  8 &  -0.4719 &  0.0000 & 0.9597 & -0.0453  \\
  10 & -0.4721 &  0.0000 & 0.9602 & -0.0452  \\
  12 & -0.4721 &  0.0000 & 0.9600 & -0.0452  \\
  14 & -0.4721 &  0.0000 & 0.9601 & -0.0452  \\
  16 & -0.4721 &  0.0000 & 0.9601 & -0.0453  \\
  18 & -0.4721 &  0.0000 & 0.9601 & -0.0453  \\
  20 & -0.4721 &  0.0000 & 0.9601 & -0.0453  \\
  \hline
\end{tabular} 
\label{tab:he6_results}
\end{table}
Our calculations are comparable with the experimental values 
of $-0.98$ MeV for the $0^+$ ground state
and $1.8-0.06i$ MeV for the $2^+$ excited state in $^6$He. Especially the 
$2^+$ excited state is well reproduced.
A splitting of $\sim 1.5$ MeV is obtained
between the $0^+$ ground state and the $2^+$ excited state, to be compared with the 
experimental value of $1.8$ MeV. The binding energy of the ${0}^+$ ground-state
may be improved by going beyond first order in 
many-body perturbation theory, as shown in for example Ref.~\cite{hko95}. 
The $2p-1h$ and $1p-2h$ diagrams may yield extra binding and improve
spin-orbit splitting for the HF single-particle states of $^4$He. In addition, it is well-known that the
two-body core-polarization contributions improve the spectroscopy of systems with two and more valence
nucleons \cite{hko95}. These findings agree also with Brueckner-Hartree-Fock calculations 
for oxygen isotopes, see for examples Ref.~\cite{bhf1990}. At first order in perturbation theory, the spectrum
is very much compressed. The agreement with experiment is partly improved with the introduction of core-polarization
contributions.

It can be concluded that the energies considered here for $^6$He, converge
with respect to the number of harmonic oscillator functions in the expansion of the nucleon-nucleon
interaction. However, this does not imply that other observables converge with the same speed,
especially those which are sensitive to the tail of the wave function.

\begin{figure}
\resizebox{8cm}{5cm}{\epsfig{file=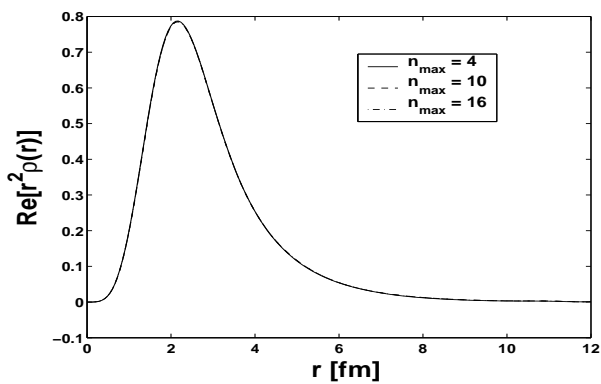}}
   \caption{Plot of the radial density of the $0^+$ ground state of $ ^6$He 
     for three different harmonic oscillator expansions with respectively $n_{\mathrm{max}} =$ 4, 10 and 16.}
\label{fig:0plus1} 
\end{figure}

\begin{figure}
\resizebox{8cm}{5cm}{\epsfig{file=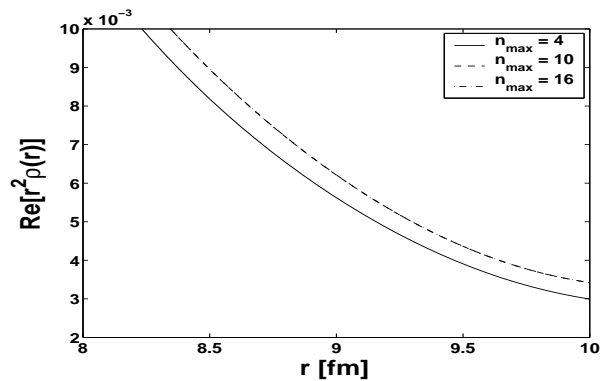}}
   \caption{Same as in Fig.~\ref{fig:0plus1}, but for the radial interval 8 $\leq$ r $\leq$ 10.}
   \label{fig:0plus2} 
\end{figure}

In order to investigate such a dependence, the single-particle radial 
density operator is considered for the $0^+_1$ ground state 
in $^6$He. The single-particle radial density operator is given by 
\begin{equation}
  \hat{\rho} = \sum_{i}^N \vert r_{i}\rangle\langle r_{i} \vert,
\end{equation}
where $N$ is the total number of valence particles ($N = 2$ for $^6$He in a $^4$He core). 
This operator measures the probability for
either particle $1$ or $2$ is to be found at the position $r_i$.
Fig.~\ref{fig:0plus1}  shows plots of the diagonal part  of $\hat{\rho}$, that is  
$\rho(r = r_1 = r_2)$ for the 
$0^+$ ground state of $^6$He, with $n_{\mathrm{max}} = 4, 10$ and 16 
in the harmonic oscillator expansion of the nucleon-nucleon interaction. 
In Fig.~\ref{fig:0plus1} there is no observable difference between 
the $n_{\mathrm{max}} = 4, 10$ and 16 results. In Fig.~\ref{fig:0plus2} we examine  
the tail of the wave function. The densities obtained
for $n_{\mathrm{max}} = 10 $ and $16$ are indistinguishable, while the
$n_{\mathrm{max}} = 4 $ displays a very small deviation from the converged results.

\subsection{ $^{18}$O results }
Tab.~\ref{tab:o18_data1} gives the convergence of the ${0_1}^+, {0_2}^+, {4_1}^+  $ 
and  $ {4_2}^+ $ state energies and Tab.~\ref{tab:o18_data2}  
 the convergence of the ${2_1}^+, {2_2}^+, {2_3}^+  $ and  $ {2_4}^+ $ of $^{18}$O, as 
the number of nodes in the harmonic oscillator expansion increases. 
All of the states converge with 
a reasonable low number of harmonic oscillator functions $n_{\mathrm{max}} \sim 10$.
Our calculation of the $0^+_1$ ground-state energy comes at $-12.23$MeV which is
very close to the experimental value of $-12.18$MeV. The calculated 
splitting between the  $0^+_1$ ground-state and the $0^+_2$  excited state 
is $\sim 3.73$MeV which is also very close to the experimental value $3.63$MeV.
We are also able to predict that the first excited resonant state is the 
$ {4_2}^+$ state coming at energy $ -1.44  -0.74i$MeV, which is in agreement with 
experiment. However, we are not able to correctly describe the splitting 
between the ${0_1}^+, {2_1}^+$ and  the ${4_1}^+$ states. In our calculations
the ${0_1}^+ $ and $ {2_1}^+$ are almost degenerate. The discrepancy with 
experimental data is expected to be reduced by going beyond first order in perturbation theory, 
including contributions such as the core-polarization diagrams to the effective interaction.
In Ref.\cite{hko95}, it was shown how the splitting of the
${0_1}^+, {2_1}^+$ and  the ${4_1}^+$ states of $^{18}$O is indeed improved by 
going to higher order in perturbation theory.
In order to improve our Gamow shell-model calculations for $^{18}$O starting from 
realistic interactions, we must include higher order diagrams in the 
perturbation series for the nucleon self-energy and the effective interaction.  This is a topic which 
will be followed up in the future. 

\begin{table}[htbp]
\caption{Convergence of the ${0_1}^+, {0_2}^+, {4_1}^+ $ and  the $ {4_2}^+ $ energies in $^{18}$O
as functions of the number of harmonic oscillator nodes in the harmonic oscillator expansion.
A model space defined by $\Lambda = 1.9\mathrm{fm}^{-1}$ was used. 
The harmonic oscillator length is fixed at $b=2$ fm. Energies are in units of MeV.}
\begin{tabular}{rrrrrrrrr}\hline
  \multicolumn{1}{c}{} & 
  \multicolumn{2}{c}{$J^\pi = {0_1}^+$} & 
  \multicolumn{2}{c}{$J^\pi = {0_2}^+ $}&
  \multicolumn{2}{c}{$J^\pi = {4_1}^+$}&
  \multicolumn{2}{c}{$J^\pi = {4_2}^+$}\\ 
  \hline
  \multicolumn{1}{c}{$n_{\mathrm{max}}$} & 
  \multicolumn{1}{c}{Re[E]} & \multicolumn{1}{c}{Im[E] }& 
  \multicolumn{1}{c}{Re[E]} & \multicolumn{1}{c}{Im[E] }&
  \multicolumn{1}{c}{Re[E]} & \multicolumn{1}{c}{Im[E] }&
  \multicolumn{1}{c}{Re[E]} & \multicolumn{1}{c}{Im[E] }\\
  \hline
  4 &  -12.225 & 0.000 & -8.438 & 0.000 & -11.0641 & 0.0000 & -1.4373 & -0.8275 \\
  6 &  -12.226 & 0.000 & -8.498 & 0.000 & -11.0907 & 0.0000 & -1.4292 & -0.7600 \\
  8 &  -12.228 & 0.000 & -8.499 & 0.000 & -11.0922 & 0.0000 & -1.4380 & -0.7405 \\
  10 & -12.229 & 0.000 & -8.499 & 0.000 & -11.0921 & 0.0000 & -1.4400 & -0.7390 \\
  12 & -12.228 & 0.000 & -8.499 & 0.000 & -11.0923 & 0.0000 & -1.4393 & -0.7401 \\
  14 & -12.228 & 0.000 & -8.499 & 0.000 & -11.0923 & 0.0000 & -1.4394 & -0.7401 \\
  16 & -12.228 & 0.000 & -8.499 & 0.000 & -11.0923 & 0.0000 & -1.4394 & -0.7401 \\
  18 & -12.228 & 0.000 & -8.499 & 0.000 & -11.0923 & 0.0000 & -1.4394  &-0.7401 \\
  20 & -12.228 & 0.000 & -8.499 & 0.000 & -11.0923 & 0.0000 & -1.4394 & -0.7401 \\
  \hline
\end{tabular} 
\label{tab:o18_data1}
\end{table}

\begin{table}[htbp]
\caption{Convergence of the ${2_1}^+, {2_2}^+, {2_3}^+ $ and  the $ {2_4}^+ $ energies in $^{18}$O
as functions of the number of harmonic oscillator nodes in the harmonic oscillator expansion.
A model space defined by $\Lambda = 1.9\mathrm{fm}^{-1}$ was used. 
The harmonic oscillator length is fixed at $b=2$ fm. Energies are in units of MeV.}
\begin{tabular}{rrrrrrrrr}\hline
  \multicolumn{1}{c}{} & 
  \multicolumn{2}{c}{$J^\pi = {2_1}^+$} & 
  \multicolumn{2}{c}{$J^\pi = {2_2}^+ $}&
  \multicolumn{2}{c}{$J^\pi = {2_3}^+$}&
  \multicolumn{2}{c}{$J^\pi = {2_4}^+$}\\ 
  \hline
  \multicolumn{1}{c}{$n_{\mathrm{max}}$} & 
  \multicolumn{1}{c}{Re[E]} & \multicolumn{1}{c}{Im[E] }& 
  \multicolumn{1}{c}{Re[E]} & \multicolumn{1}{c}{Im[E] }&
  \multicolumn{1}{c}{Re[E]} & \multicolumn{1}{c}{Im[E] }&
  \multicolumn{1}{c}{Re[E]} & \multicolumn{1}{c}{Im[E] }\\
  \hline
  4 &    -12.1398 & 0.0000 & -10.0488 & 0.0000 & -0.0772 &-1.4465 & 1.1632 & -1.4478 \\
  6 &    -12.1465 & 0.0000 & -10.0830 & 0.0000 & -0.1321 &-1.3038 & 1.1628 & -1.5059 \\
  8 &    -12.1452 & 0.0000 & -10.0853 & 0.0000 & -0.1539 &-1.2929 & 1.1836 & -1.5346 \\
  10 &   -12.1450 & 0.0000 & -10.0857 & 0.0000 & -0.1595 &-1.2922 & 1.1807 & -1.5331 \\
  12 &   -12.1453 & 0.0000 & -10.0858 & 0.0000 & -0.1570 &-1.2938 & 1.1820 & -1.5343 \\
  14 &   -12.1453 & 0.0000 & -10.0858 & 0.0000 & -0.1571 &-1.2938 & 1.1821 & -1.5342 \\
  16 &   -12.1453 & 0.0000 & -10.0858 & 0.0000 & -0.1573 &-1.2936 & 1.1822 & -1.5342 \\
  18 &   -12.1453 & 0.0000 & -10.0858 & 0.0000 & -0.1573 &-1.2936 & 1.1822 & -1.5342 \\
  20 &   -12.1453 & 0.0000 & -10.0858 & 0.0000 & -0.1573 &-1.2936 & 1.1822 & -1.5342 \\               
  \hline
\end{tabular} 
\label{tab:o18_data2}
\end{table}


\section{Equivalence between position and momentum space 
representations}\label{sec:app2}

While the momentum representation of one-body Gamow states has been used in \cite{hagen1,hagen2} and here,
the Gamow shell-model 
was first introduced  employing a  position space representation, see for example Refs.~\cite{witek1,roberto}.
A momentum space representation has however normally been preferred in constructions of 
effective interactions based on realistic nucleon-nucleon interaction models. There are several reasons for this.
Realistic interactions are usually derived in $k$-space, so that plane wave expansions
are the most natural bases.
Moreover, in connection with Gamow shell-model calculations, the momentum space representation was meant to
lead to a faster convergence with the number of discretized 
scattering states, as Gamow wave functions in momentum space are usually more localized compared to those 
in the coordinate representation. 
Moreover, the momentum space representation of the one-body Schr\"odinger equation is an integral 
equation in the general case, contrary to its integro-differential form in $r$-space,
known to be much more difficult to solve. The necessity of imposing 
the asymptotics of the Gamow state in $r$-space was also thought to give rise to a slower convergence with the 
number of discretized scattering states. Finally, the use of complex scaling \cite{gyarmati1} to calculate two-body matrix elements 
in the position representation leads to extremely slow calculations 
and cannot even regularize a large class of
infinite matrix elements occurring in long-range interactions. It was this last nuisance which
motivated the use of surface-peaked interactions in position space calculations, namely the Surface Delta
Interaction (SDI) \cite{witek1,witek2} and the Surface Gaussian Interaction (SGI) \cite{michel1,michel2}.
One will see however that both representations are in fact equivalent
theoretically but also numerically, where the computational cost to obtain a given precision is 
comparable in both cases. The main point is that the possibility to use the harmonic oscillator 
expansion method of Sec.~\ref{sec:vosc}
removes all the problems previously encountered with both representations.

In order to obtain a self-consistent Gamow Hartree-Fock basis in position space, one has to
solve the one-body integro-differential Schr\"odinger equation
\begin{eqnarray}
\frac{\hbar^2}{m_{\mathrm{eff}}} \varphi_{nlj}'' (r) + 
\int_0^{+ \infty} dr'\:  {\cal V}_{HF}(jlrr') \varphi_{nlj} (r') = E_{nl} \varphi_{nlj} (r),
\end{eqnarray}
where $ {\cal V}_{HF}(jlrr') $ is the self-consistent HF potential given in 
Eq.~(\ref{eq:hf_coord}). The wave function
$|\varphi_{nlj} \rangle$ has to exhibit a  pure outgoing wave function
behavior for bound and resonant states for $r \rightarrow +\infty$, whereas it has 
both incoming and outgoing components if it is a scattering state.
In the last case, $n$ must be understood as representing its wave number $k$.
Even though it is an integro-differential equation, it can be solved with 
standard methods with the use of locally equivalent potentials \cite{vaut1},
so that its integration has the same complexity as differential equations occurring with purely local potentials. 
This generates a self-consistent
procedure, as the locally equivalent potential depends on the state that it generates. 
This is of no importance in practical situations, as 
the integrated potential will be non-local in $r$-space only if it is a 
HF potential generated by a finite-range interaction.
As HF potentials have to be solved self-consistently, no numerical overhead can occur. 
The slow convergence noted in Ref.~\cite{witek1}
was due only to the use of the trapezoidal rule
to discretized the non-resonant continuum. With the use of the Gauss-Legendre integration, 
as performed in $k$-space calculations from the beginning,
results have improved dramatically in $r$-space calculations, reaching $k$-space calculations quality \cite{michel5}.

In fact, the fundamental difference between the $r$ and the $k$ representations for Gamow shell-model 
applications lies in their different discretization schemes.
In $k$-space, it is the Bessel completeness relation of Eq.~(\ref{eq:bessel_complete})
which is discretized. 
The $|\psi_{nlj} \rangle$ states are then obtained by diagonalization of ${\cal V}_{HF}(jlkk')$
in the discretized Fourier-Bessel basis space. They will be denoted as $|\psi_{nl} \rangle^{D_k}$. 
In $r$-space, it is the completeness relation 
spanned by the $|\varphi_{nl} \rangle$ states themselves which is discretized. Indeed, one has : 
\begin{eqnarray}
&& \sum_{n \in (b,d)} |\varphi_{nl} \rangle \langle \varphi_{nl}| + \int_{L_+} |\varphi_{kl} \rangle \langle \varphi_{kl}| \; dk = \mathbbm{1} \label{exactcomp} \\
&& \sum_{i=1}^{N}  w_n |\varphi_{nl} \rangle \langle \varphi_{nl}| \simeq \mathbbm{1}, \label{appcomp}
\end{eqnarray}
where $n \in (b,d)$ means that one sums over all bound $(b)$ and decaying $(d)$ states above the contour $L_+$.
The first completeness relation, exact, becomes the second discretized completeness relation, approximate, where $w_n$
is 1 for bound and resonant states, and the Gauss-Legendre weight for scattering states.
As a consequence, the $|\varphi_{nl} \rangle$ states of $r$-space, that we 
label $|\varphi_{nl} \rangle^{D_r}$ are exact up to numerical precision, 
since they come from a direct integration of the Schr\"odinger equation. 
Approximations arise only from their discrete and finite number in Eq.~(\ref{appcomp}). 
The corresponding $|\psi_{nl} \rangle^{D_k}$ states have to be approximate, as they are generated 
by a finite number of Bessel basis states,
meaning that there are not enough frequencies to expand them exactly.
As a consequence, one has $|\psi_{nl} \rangle^{D_k} \rightarrow |\psi_{nl} \rangle^{D_r}$ 
only at the continuum limit, that is  for $N \rightarrow +\infty$.

This difference may appear for observables that depend on large values of $r$ or $k$, 
such as particle densities at large $r$ or momentum densities
at large values of $k$. But this is of no importance as discretization effects  
become preponderant 
in these regions, and may thereby most likely  lead to numerically
unstable results.

Consequently, both representations can be used in shell-model problems 
without any loss of precision. The remaining question of the method to handle
two-body matrix elements in purely numerical bases has been answered in Sec.~\ref{sec:vosc}
with the use of harmonic oscillator expansions. One has seen in the latter section that the only numerical calculations
involving Gamow states are the overlaps between Gamow and harmonic oscillator states of Eq.~(\ref{eq:sp_overlap}),
which are obviously fast numerically. The Gaussian decrease of harmonic oscillator states in momentum or position
representation allows a very accurate implementation of overlaps in both representations.
Hence, the implementation of the Gamow shell-model matrix becomes very similar from one representation to another.

The Coulomb interaction, not considered in this paper, 
may however generate difficulties in the momentum representation.
Its infinite range character can indeed be treated exactly in the $r$-representation 
at the basis level through the use of
Coulomb wave functions, whereas approximations have to be 
performed in $k$-space calculation as the Fourier-Bessel transform
of the Coulomb interaction does not exist. 
The use of the harmonic oscillator expansion method may be indeed slowly converging for its low multipoles.
This question will have to be answered with Gamow shell-model calculations of nuclei close to the proton drip-line.

\section{Conclusion and future perspectives}
\label{sec:conclusion}
In this paper we have presented  a calculational algorithm which can be used  to obtain 
a self-consistent single-particle
basis in the complex energy plane, starting from a 
renormalized and realistic nucleon-nucleon interaction. In this work we
used the simplest possible approximation to the nucleon self-energy, 
including the Hartree-Fock diagram only in order to demonstrate the feasibility of our method. 
With our approach we studied the single-particle 
spectra of $^4$He and $^{16}$O. For $^4$He we found that both 
the $p_{3/2}$ and the $p_{1/2}$ states appeared as resonances.
Their widths are in fair agreement with experiment. 
For $^{16}$O we found the hole states to be  largely overbound, 
while the $s_{1/2}$ and $d_{5/2}$ particle states
agree well with the  experimental values. The $d_{3/2}$ states comes out as a resonance, in agreement with experiment 
although our width is larger
than the experimental value.  
Higher order corrections such as $2p-1h$ and $1h-2p$ contributions may improve the agreement with experiment, in particular the 
spin-orbit splittings for the hole states, 
and will be included in future self-energy  calculations.

With the Gamow Hartree-Fock single-particle basis, derived 
from realistic interactions, the problem of representing the nucleon-nucleon 
interaction in the derived basis comes to the fore. As the nucleon-nucleon interaction is typically given 
in momentum space, a transformation from the relative and center of mass frame to the laboratory frame involves 
mathematical functions which are not easy to continue analytically in the complex $k$-plane. 
To that end, we investigated whether a method based on an expansion of the nucleon-nucleon interaction 
as function of a  finite set of harmonic oscillator functions could be a promising route. 
This expansion allows for 
a straightforward calculation of matrix elements in the laboratory frame for any Gamow basis.
The harmonic oscillator functions are indeed very flexible  in  both position and momentum space, 
and the analytic continuation of the two-body interaction in the complex plane
turns out to be very easy to implement from both a theoretical and  a numerical point of view.
With this method, we  have shown  that for the example of the
$^6$He and $^{18}$O nuclei, all states converge with a low number of harmonic oscillator functions 
in the expansion of the interaction. 
This method offers also  
a practical way of calculating higher order diagrams in many-body perturbation theory.
There are diagrams which enter for example the definition
of the self-energy and two or three-body effective interactions. 
In particular, this method provides a solution to the problem of non-orthogonal 
intermediate particle states, a problem which arises when one uses the vector brackets. Utilizing 
the harmonic oscillator expansion of the
interaction, we ensure that all intermediate states are orthogonal 
in all diagrams beyond first order in many-body perturbation theory. However, the renormalization
of the Coulomb interaction and the two-body center of mass contribution need further considerations.

The procedure outlined in this work allows for several interesting applications related to 
the study of weakly bound and unbound nuclei along the driplines.  
Of particular interest is the possibility to apply our approach within the framework of the 
coupled-cluster method, see for example 
Refs.~\cite{coester,coesterkummel,cizek,palduscizek,cc2,cc3}.
The coupled-cluster approach
is a promising
candidate for the development of practical methods for
fully microscopic {\em ab initio} studies of nuclei. 
The coupled-cluster methods are capable of providing a precise description
of many-particle correlation effects at relatively low computer costs,
when compared to shell-model or configuration interaction techniques aimed at
similar accuracies.
In approaches 
such as the coupled cluster, 
an extension to the complex energy plane is in principle 
possible. 
Coupled-cluster calculations starting with a HF basis in the complex plane,
along with the inclusion of realistic interactions, are planned in near future. 
Then, it might be possible
to perform coupled-cluster calculations of states with a 
multi-particle resonant structure starting with a realistic 
interaction. 
 
\section*{Acknowledgments}
Discussions with Jimmy Rotureau are gratefully acknowledged.
This work was supported in part by the U.S.~Department of Energy
under Contracts Nos.~DE-FG02-96ER40963 (University of Tennessee),
DE-FG05-87ER40361 (Joint Institute for Heavy Ion
Research), and the Research Council of Norway
(Supercomputing grant NN2977K). 
Oak Ridge National Laboratory is managed by UT-Battelle for the
U.S. Department of Energy under Contract No.~DE-AC05-00OR22725.


\end{document}